\documentclass[fleqn,usenatbib]{mnras}

\usepackage{newtxtext,newtxmath}

\usepackage[T1]{fontenc}
\usepackage{ae,aecompl}

\usepackage{graphicx}	
\usepackage{amsmath}	
\usepackage{amssymb}	

\usepackage{caption}
\usepackage{subcaption}
\title[Plerion model of sGRB plateaux]{Plerion model of the X-ray plateau in short gamma-ray bursts}

\author[L. C. Strang and A. Melatos]{
L. C. Strang,$^{1, 2}$\thanks{E-mail: lstrang@student.unimelb.edu.au}
A. Melatos,$^{1, 2}$
\\
$^{1}$School of Physics, University of Melbourne, Parkville, VIC 3010 Australia\\
$^{2}$ Australian Research Council Centre of Excellence for Gravitational Wave Discovery (OzGrav), \\
\,  University of Melbourne, Parkville, VIC 3010, Australia \\
}

\date{Accepted 2019 June 5. Received 2019 May 28; in original form 2019 February 3}

\pubyear{2019}

\begin{document}
\label{firstpage}
\pagerange{\pageref{firstpage}--\pageref{lastpage}}
\maketitle

\begin{abstract}
  Many short gamma-ray bursts (sGRBs) exhibit a prolonged plateau in the X-ray light curve following the main burst. It is shown that an X-ray plateau at the observed luminosity emerges naturally from a plerion-like model of the sGRB remnant, in which the magnetized, relativistic wind of a millisecond magnetar injects shock-accelerated electrons into a cavity confined by the sGRB blast wave.
  A geometry-dependent fraction of the plerionic radiation is also intercepted and reprocessed by the optically thick merger ejecta.
  The relative contributions of the plerion and ejecta to the composite X-ray light curve are estimated approximately with the aid of established ejecta models.
  The plerionic component of the electron energy spectrum is evolved under the action of time-dependent, power-law injection and adiabatic and synchrotron cooling in order to calculate the X-ray light curve analytically. The model yields an anti-correlation between the luminosity and duration of the plateau as well as a sudden cut-off in the X-ray flux, if the decelerating magnetar collapses to form a black hole. Both features are broadly consistent with the data and can be related to the surface magnetic field of the magnetar and its angular velocity at birth. The analogy with core-collapse supernova remnants is discussed briefly.
\end{abstract}




\section{Introduction}

Many short gamma-ray bursts (sGRBs) are observed to have prolonged X-ray afterglows, which are long-lived in comparison to the main burst.
The lifetimes and evolution of these afterglows are often interpreted as evidence for some long-lived central engine which survives the burst \citep{Metzger2008,Nousek2006,Zhang2006}.
The recent coincident detection of GW170817 \citep{Abbott2017a,Abbott2017b} by the Advanced Laser Interferometer Gravitational-wave Observatory (LIGO) and GRB 170817a \citep{Goldstein2017,Savchenko2017,Zhang2017} by multiple telescopes confirms that binary neutron star mergers are the progenitors of some sGRBs.
However, the detailed evolution of post-merger remnants remains uncertain.
There is a plethora of models predicting a variety of different observational signals following a collision event  \citep{Mooley2018,Sari1999,Nathanail2018,Metzger2014,Lyutikov2017,Veres2018}.
The remnant of the merger may be a stable neutron star, a black hole, or an unstable magnetar which collapses to a black hole after it loses sufficient rotational energy \citep{Cook1994b,Cook1994a}.
The outcome is dictated largely by the equation of state \citep{Lattimer2001}.

Of the sGRBs with extended X-ray afterglows, many display common features including a plateau and decay phase, which have been termed  `canonical' by various authors \citep{Zhang2006,Nousek2006}.
Canonical X-ray light curves display an initial steep decay followed by a plateau, where the luminosity is approximately constant.
The plateau phase lasts for 10 -- 10$^5$ s and typically starts 1 -- $10^3$ s after the prompt emission.
After the plateau, the light curve again decays but more slowly than initially.
The luminosity of the plateau is anti-correlated with its duration in both short and long GRBs \citep{Dainotti2010}.
In a sample of 43 sGRBs from the \textit{Swift} telescope, \citet{Rowlinson2013} identified 37 with an  X-ray afterglow, 22 of whose light curves are well described by a broken power law with three components, including a plateau.

A millisecond magnetar model explains several observational features of canonical X-ray light curves \citep{Dai1998,Zhang2001,Fan2006,Lasky2017}.
Some existing analyses of the millisecond magnetar model concentrate on explaining the temporal evolution of the X-ray flux \citep[e.g.][]{Lasky2017}. 
Other models have considered the temporal and spectral evolution of a millisecond magnetar shrouded by merger ejecta \citep{Yu2013,Metzger2014,Siegel2016,Siegel2016b}.

In this paper, we generalize previous work to incorporate the spectral evolution of the source.
In the first instance, as a simple idealization, we adopt the framework of the classic plerion model of supernova remnant evolution \citep{Pacini1973}, thereby concentrating initially on the geometry-dependent fraction of the radiation that is not intercepted and reprocessed by the optically thick merger ejecta.
Our goal is to test whether particle injection from a millisecond magnetar, together with adiabatic and synchrotron cooling, can deliver the observed luminosity at the correct (X-ray) frequencies at the correct time.
We then estimate approximately the non-thermal radiation from the ejecta using established models in the literature \citep{Yu2013,Metzger2014,Siegel2016,Siegel2016b} and compute a composite light curve.

The paper is structured as follows. In Section \ref{sec:plerionmodel}, we introduce the plerion model and describe the analogy with sGRB remnants.
We calculate the spectral evolution of the plerionic component of the remnant in Section \ref{sec:spectralevolution} and present the resulting light curves in Section \ref{sec:xray}.
The contribution of the merger ejecta is estimated in Section \ref{sec:ejecta}.
Practical analytic formulas for key observable quantities are developed for useful special cases in the Appendix.

\section{Plerion model}
\label{sec:plerionmodel}

We consider a simple model for X-ray emission through synchrotron radiation from relativistic electrons injected into a shocked bubble by the post-merger central engine, taken to be a stable or unstable neutron star.
This scenario shares some similarities with the classic plerion model of supernova remnants \citep{Pacini1973}.
The model presented here consists of two key components: the blast wave, and the relativistic electrons injected by the pulsar wind, which are described in detail in Sections \ref{sec:bw} -- \ref{sec:magfield}.

As the result of a merger, the star may exceed the canonical neutron star mass, $M_* = 1.4 M_\odot$, if it is supported by uniform rotation \citep{Rezzolla2018}.
\citet{Rezzolla2018} reported a maximum mass of $M_* \leq 2.59 M_\odot$, and \citet{Ai2018} suggested the GW170817 remnant may have a mass of $M_* \approx 2.57 M_\odot$ using similar methods.
The radius $R_*$ of remnants of neutron star merger events is less well constrained.
Recent work \citep[e.g.][]{Bauswein2017,2016ARAA..54..401O} suggests neutron stars in general may have a radius satisfying $9.6 \text{ km} < R_* < 11.5 \text{ km}$ depending on the equation of state. 
Throughout this work, we take standard neutron star parameters, i.e. $M_* = 1.4 M_\odot$ and $R_* = 10^4$ m, but it is easy to recompute the results across the ranges presented above. 

A neutron star merger also produces $10^{-3} M_{\odot}$ to $10^{-2} M_\odot$ of heavy ejecta material \citep{Rosswog1999}.
In the specific case of GW170817, optical and near-infrared observations suggest that up to $0.03 M_\odot$ of \textit{r}-process debris fuelled the associated kilonova \citep{Tanaka2017}. 
We estimate briefly the contribution of the merger ejecta to the light curve in Section \ref{sec:ejecta}.

\subsection{Blast wave}
\label{sec:bw}
Immediately following the merger, a blast wave of radius $r_b(t) \approx c t$ expands into the interstellar medium (ISM).
We treat the blast wave as a relativistic point explosion with energy $E_b$ expanding into a uniform medium (mass density $\rho_{ \text{ISM} }$), described by the self-similar solution developed by \citet{Blandford1976}.
The solution states that the energy behind the shock is concentrated in the expanding shell of outward moving particles.
Little energy is left at the site of the explosion, i.e. at $r \ll r_b(t)$.
The number density of the shocked electrons scales as $n \propto (1-r/ct)^{-7/4}$ and their Lorentz factor varies as $\gamma_e \propto (1-r/ct)^{-1}$.
Henceforth we make the simplifying approximation that all the electrons are contained in a thin shell at radius $r_b$.

\subsection{Pulsar spin-down luminosity}
\label{sec:spdlum}
We assume that the merger leaves behind a rapidly rotating neutron star, which brakes electromagnetically as it emits a magnetized leptonic wind.
As the pulsar spins down, its rotational energy is converted into a mixture of mechanical and electromagnetic energy flux.

Initially the neutron star rotates near the centrifugal breakup frequency \citep{Cook1994b} with angular velocity $\Omega(t=0) = \Omega_0 \approx 6\times 10^3$ rad s$^{-1}$.
The braking law can be written as 
\begin{equation}
 \frac{d\Omega}{dt} = - k\Omega^n.
 \label{eqn:dOmegadt}
\end{equation}
Here $k$ is a proportionality constant related to the (polar) surface magnetic field $B_0$, and $n$ is the braking index.
The precise value of $n$ for realistic models remains uncertain; it is measured in the range $2 \leq n \leq 7$ in ordinary radio pulsars \citep{Melatos1997,Archibald2016} and falls close to the vacuum dipole value $n = 3$ in recent fits to X-ray plateaux in sGRBs \citep{Lasky2017}.
Taking $n=3$ for definiteness, the solution to (\ref{eqn:dOmegadt}) is
\begin{equation}
\Omega(t) = \Omega_0 \left( 1+ \frac{t}{\tau} \right)^{-1/2}
\label{eqn:spindownfreq}
\end{equation}
where 
\begin{equation}
\tau = \frac{3 c^3 \mu_0 I}{4\pi\Omega_0^2 R_*^6 B_0^2}
\label{eqn:spindowntime}
\end{equation}
is the characteristic spin-down time-scale and $I$ is the moment of inertia \citep{Zhang2001}. 
Equations (\ref{eqn:spindownfreq}) and (\ref{eqn:spindowntime}) neglect gravitational radiation reaction.

The spin-down luminosity of the pulsar $I \Omega \dot{\Omega}$ is given by
\begin{equation}
L(t) = L_0 \left( 1+ \frac{t}{\tau}\right)^{-2}
\label{eqn:spindownlum}
\end{equation}
where
\begin{equation}
  L_0 = \frac{I \Omega_0^2}{2\tau}
  \label{eqn:spindownluminit}
\end{equation}
is the initial spin-down luminosity.

\subsection{Particle injection radius}
\label{sec:pinjr}
We suppose shock-accelerated electrons are injected evenly into a spherical shell with $r_{\text{inj}} \leq r \leq r_b$.
In reality, the system is more complicated and almost certainly anisotropic, but sGRBs are observed as point sources, so we average over the anisotropy here. 
The electrons are injected at a rate $\dot{N}_{\text{inj}} \propto L(t)$ at highly relativistic speeds.
The mean Lorentz factor $\gamma_e$ of the injected electrons is much greater than the Lorentz factor of the blast wave.

The injection radius $r_{\rm inj}$ is located, where the static pressure of the accumulated population of accelerated electrons, $P_{\rm stat}(r_{\rm inj})$, balances the ram pressure $P_{\rm ram}(r_{\rm inj})$ of the pulsar wind, i.e. $P_{\rm stat}(r_{\rm inj})=P_{\rm ram}(r_{\rm inj})$.
For a relativistic outflow we have
\begin{equation}
P_{ \text{ram} } = \frac{\dot{N}_{\text{inj}} \langle E \rangle}{4\pi r_{ \text{inj} }^2 c},
  \label{eqn:pram}
\end{equation}
where $\langle E \rangle = \langle \gamma_e \rangle m_e c^2 \gg m_e c^2$ is the mean energy of the injected electrons.
The static pressure arises from the electrons injected up to time $t$ and the energy $E_{\text{shell} }$ left over from the initial blast wave, where the energy of the blast wave is assumed to be negligible except inside the shell.
The static pressure is then
\begin{equation}
P_{\text{stat} } = V(r_{\text{inj}})^{-1} \left[ E_{\text{shell} } + \int_0^t dt' L(t') \right] ,
\label{eqn:pstat}
\end{equation}
where $V(r_{\text{inj}})$ is the volume of the shell.

\begin{figure}
  \centering
  \includegraphics[width=\linewidth]{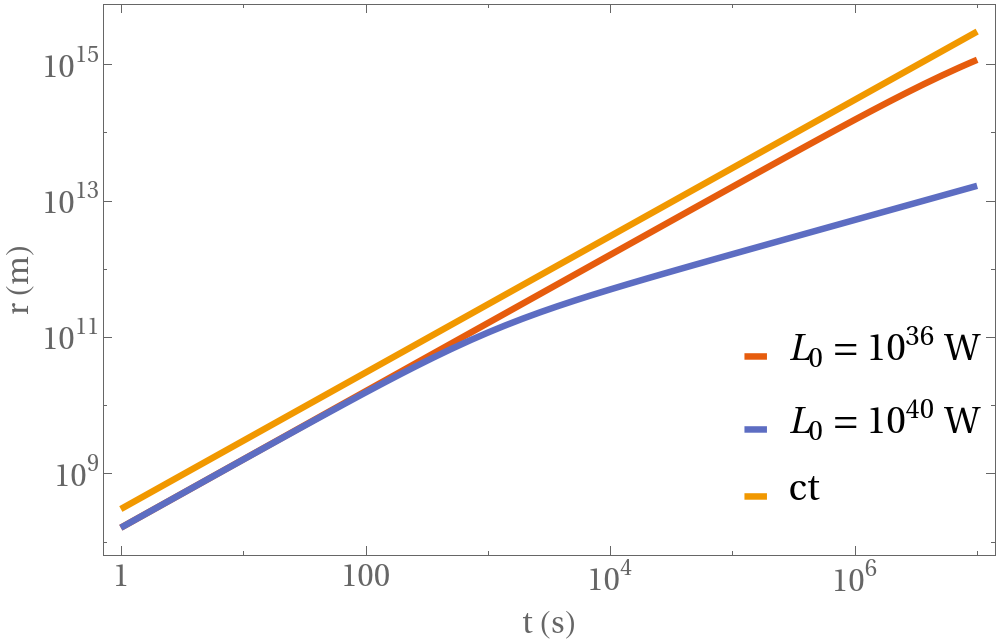}
  \caption{\label{fig:rinj} Injection radius $r_{\text{inj}}$ (m) as a function of time (s), obtained by solving $P_{\text{stat}}(r_{\text{inj}}) = P_{\text{ram}}(r_{\text{inj}})$ for $L_0 = 10^{36} \, {\rm W }$ (red curve) and $L_0 = 10^{40} \, {\rm W}$. The orange curve shows $r_{\rm inj} = c t$ for comparison.}
\end{figure}

The injection surface moves almost at the speed of light for $t < \tau$ and at a high fraction of $c$ for $t > \tau$ (Figure \ref{fig:rinj}).
The injection radius turns around and travels inward as a reverse shock, i.e. $\dot{r}_{\rm inj} < 0$, for $t \gtrsim 10^8\,{\rm s}$ for $L_0 \geq 10^{40}\,{\rm W}$.
The turn-around occurs earlier for higher $L_0$, because the static pressure (\ref{eqn:pstat}) grows faster.
The turn-around may occur before or after X-ray observations of the afterglow cease.

\subsection{Magnetic field}
\label{sec:magfield}

The spin-down luminosity (\ref{eqn:spindownlum}) is transported into the plerion in a relativistic magnetized wind.
Inside the light cylinder $r_L(t) = c/\Omega(t)$, the magnetic field is approximately dipolar and falls off as $r^{-3}$.
Beyond $r_L(t)$, the magnetic field is approximately monopolar and falls off as $r^{-1}$.
The magnetic field is then
\begin{equation}
B(r) = B_0 \begin{cases}
  \left( r/R_* \right)^{-3} & r < r_L(t)\\
  \left[ r_L(t)/R_*\right]^{-3} \left[ r/r_L(t)  \right]^{-1} & r \geq r_L(t)
\end{cases}
\label{eqn:magfieldradial}
\end{equation}
As the bubble of electrons  expands at speed $\dot{r}_b \approx c \gg \dot{ r }_L$, we have $r_b(t) \gg r_L(t)$ for all $t$.
Hence the magnetic field at the thin expanding shell takes the form
\begin{equation}
B(t) = \frac{B_0 R_*^3 \Omega_0^2}{c^3 t (1+t/\tau)}
\label{eqn:magfieldt}
\end{equation}
and is a function of time $t$.
For simplicity, we take the magnetic field to be uniform in the volume $r_{\rm inj} \leq r \leq r_{\rm out}$.
\section{Spectral evolution}
\label{sec:spectralevolution}
In this section, we calculate the evolution of the electron energy distribution $N(E,t)$ under the simultaneous actions of cooling and injection.
This involves solving the partial differential equation
\begin{equation}
    \frac{\partial N(E,t)}{\partial t}  = \frac{\partial  }{\partial E} \left[ \left.\frac{dE}{dt}\right|_{\text{ cool }} N(E,t)\right] + \dot{N}_{\text{inj}}(E,t)
\label{eqn:pdegen}
\end{equation}
where
\begin{equation}
 \left. \frac{dE}{dt}\right|_{\text{cool}} = \left. \frac{dE}{dt}\right|_{\text{syn}} + \left. \frac{dE}{dt}\right|_{\text{ad}}
\end{equation}
is the cooling rate as a function of energy and time due to synchrotron (syn) and adiabatic (ad) cooling.
The total injection rate (electrons per unit time) is
\begin{equation}
\dot{N}_{\text{inj}}(t) = \int dE \dot{N}_{\text{inj}}(E,t);
\label{eqn:ndott}
\end{equation}
see Section \ref{sec:injection} for an explicit expression.
We neglect radial variation in $N(E,t)$ for simplicity, and because sGRBs are observed as point sources.
A more detailed model would (for example) solve for the electron motion from the point of injection throughout the region $r_{\text{inj}}\leq r\leq r_b$.

In Appendix \ref{sec:sol}, we show how to solve (\ref{eqn:pdegen}) for various forms of $\dot{N}_{\text{inj}}(E,t)$ and $B(t)$.
In some cases, e.g. $B(t) = $ constant, this can be done analytically via a Green's function approach.

\subsection{Injected spectrum}
\label{sec:injection}
We assume that the injection rate is proportional to the spin-down luminosity of the pulsar.
Let $\sigma$ be the ratio of Poynting flux to kinetic energy flux.
In plerions like the Crab, one finds $\sigma \approx 10^{-3}$ \citep[e.g.][]{Kennel1984,Melatos1996,Bogovalov2008}.
If we assume that the injected electrons are distributed in energy as a power law $ \propto E^{-a}$ for $E_{- 0} \leq E \leq E_{+0}$, we obtain
\begin{equation}
\dot{N}_{\text{inj}}(E,t) = \frac{L(t) (a-2) E^{-a} }{(1+\sigma) \left( E_{-0}^{2-a} - E_{+0}^{2-a} \right)  }.
\end{equation}
\subsection{Cooling}
\label{sec:cooling}
Once injected, the accelerated electrons lose energy via synchrotron radiation at the rate
\begin{equation}
\left.\frac{dE}{dt}\right|_{\text{syn}} =  - c_s E^2 B(t)^2
\label{eqn:synchdef}
\end{equation}
with $c_s = 4\sigma_T c/6\mu_0(m_ec^2)^2$, where $\sigma_T$ is the Thomson cross-section.

As the electron bubble expands, the electrons cool adiabatically.
For $r_b \approx ct$, this contributes an additional term
\begin{equation}
\left.\frac{dE}{dt}\right|_{\text{ad}} = - \frac{E}{t}
\label{eqn:cooladiab}
\end{equation}
Adiabatic losses are significant at low energies and/or late times, where the condition
\begin{equation}
 (E t )^{-1} \gtrsim c_s B(t)^2
\end{equation}
is satisfied, corresponding to
\begin{equation}
 \left( \frac{E}{m_e c^2} \right)^{-1}
 \left( \frac{t}{\tau} \right)
 \left( 1 + \frac{t}{\tau} \right)^2
 \gtrsim
 2.0 \times 10^{-3}
 \left( \frac{B_0}{10^{10}\,{\rm T}} \right)^4
 \left( \frac{\Omega_0}{10^3\,{\rm rad\,s^{-1}}} \right)^6~.
\end{equation}

\subsection{Energy range}
\label{sec:energy range}
The energy range spanned by the electrons in the shell is governed by the injection history (e.g. impulsive or ongoing), the instantaneous energy range at injection, and the subsequent energy evolution.
These factors are analysed in detail in the Appendix.
For the special illustrative case of a constant magnetic field $B(t) = B$, the energy $E(t)$ of an electron injected with energy $E_i$ at time $t_i$ evolves according to
  \begin{equation}
E(t;t_i, E_i) = \left[  E_i^{-1} + c_s B^2 (t-t_i)  \right]^{-1}
\label{eqn:econst}
  \end{equation}
  for $t \geq t_i$.
  Hence if the injection is impulsive (at $t = t_i$ only), the energy range at $t \geq t_i$ is given by $E(t;t_i, E_{-0}) \leq E \leq E(t; t_i, E_{+0})$. 
  If the injection is ongoing and constant, the range is $E(t;t_i = 0, E_i = E_{-0}) \leq E \leq E_{+0}$.

  For an expanding bubble (including adiabatic cooling) we find
  \begin{equation}
E(t;t_i, E_i) = \left[ \frac{t}{E_i t_i} + c_s t \int_{t_i}^t dt' \frac{B(t')^2}{t'} \right]^{-1}.
\label{eqn:evar}
  \end{equation}
For ongoing, constant injection, the energy range is $E(t; t_i = 0, E_i = E_{-0}) \leq E \leq E_{+0}$.

\section{X-ray light curve}
\label{sec:xray}
In this section, we present the X-ray light curves predicted by the model in Sections \ref{sec:plerionmodel} and \ref{sec:spectralevolution}.
The light curves are calculated by solving (\ref{eqn:pdegen}) for $N(E,t)$ as described in the Appendix, multiplying by $(dE/dt)_{\text{syn}}$ and integrating over the relevant energy range.
We assume for simplicity that the electrons radiate at their characteristic frequency,
\begin{equation}
\nu_c = \frac{3}{2} \left(\frac{E}{m_e c^2}\right) ^2 \frac{e B(t)}{2\pi m_e}
\label{eqn:charfreq}
\end{equation}
where $e$ is the electric charge. 
The error introduced by the approximation (\ref{eqn:charfreq}) is small compared to other uncertainties in the model and can be fixed easily, when future data warrant.

Unless otherwise stated, we consider injection with $B_0 = 10^{11}$ T, $\Omega_0 = 10^3$ rad s$^{-1}$, $E_{-0} = 8\times 10^{-6}$ J, and $E_{+0} = 8\times 10^{-8}$ J.
These values are compatible with previous work in the millisecond magnetar context \citep{Lasky2017,Rowlinson2013}.
It is worth noting that, while these injection parameters affect the brightness and duration of the plateau, they do not affect its general shape.
We thus lose little predictive power by specializing to these values.

Given $N(E,t)$, the bolometric luminosity is
\begin{equation}
L_{\text{syn}}(t) = \int dE c_s B(t)^2 E^2 N(E,t),
\label{eqn:lumbol}
\end{equation}
integrated across the energy range discussed in Section \ref{sec:energy range}.
To obtain the X-ray luminosity $L_X$, we perform the same calculation, restricting ourselves to the observing band of the \textit{Swift} satellite, $0.3-10 {\rm \,  keV}$ \citep{Gehrels2004}, i.e. $7.25\times 10^{16} \leq \nu_c/({\rm Hz }) \leq 2.42\times 10^{18}$.
We find that an X-ray plateau emerges for both the full model including adiabatic expansion, and the simplified model with a constant magnetic field. 
The observable properties of the plateau, such as its flux, duration, and shape, depend on the relative importance of synchrotron and adiabatic cooling, the evolution of $B(t)$, whether or not the remnant collapses to a black hole, and the remnant properties at birth (e.g. $\Omega_0$, $B_0$).
These factors are discussed in Sections \ref{sec:flux} -- \ref{sec:birth}.
The merger ejecta intercept and reprocess a geometry-dependent fraction of the plerionic emission \citep{Yu2013,Metzger2014,Siegel2016}, as discussed in Section \ref{sec:ejecta}.

\subsection{Plateau flux}
\label{sec:flux}
\begin{figure}
  \centering
  \includegraphics[width=\linewidth]{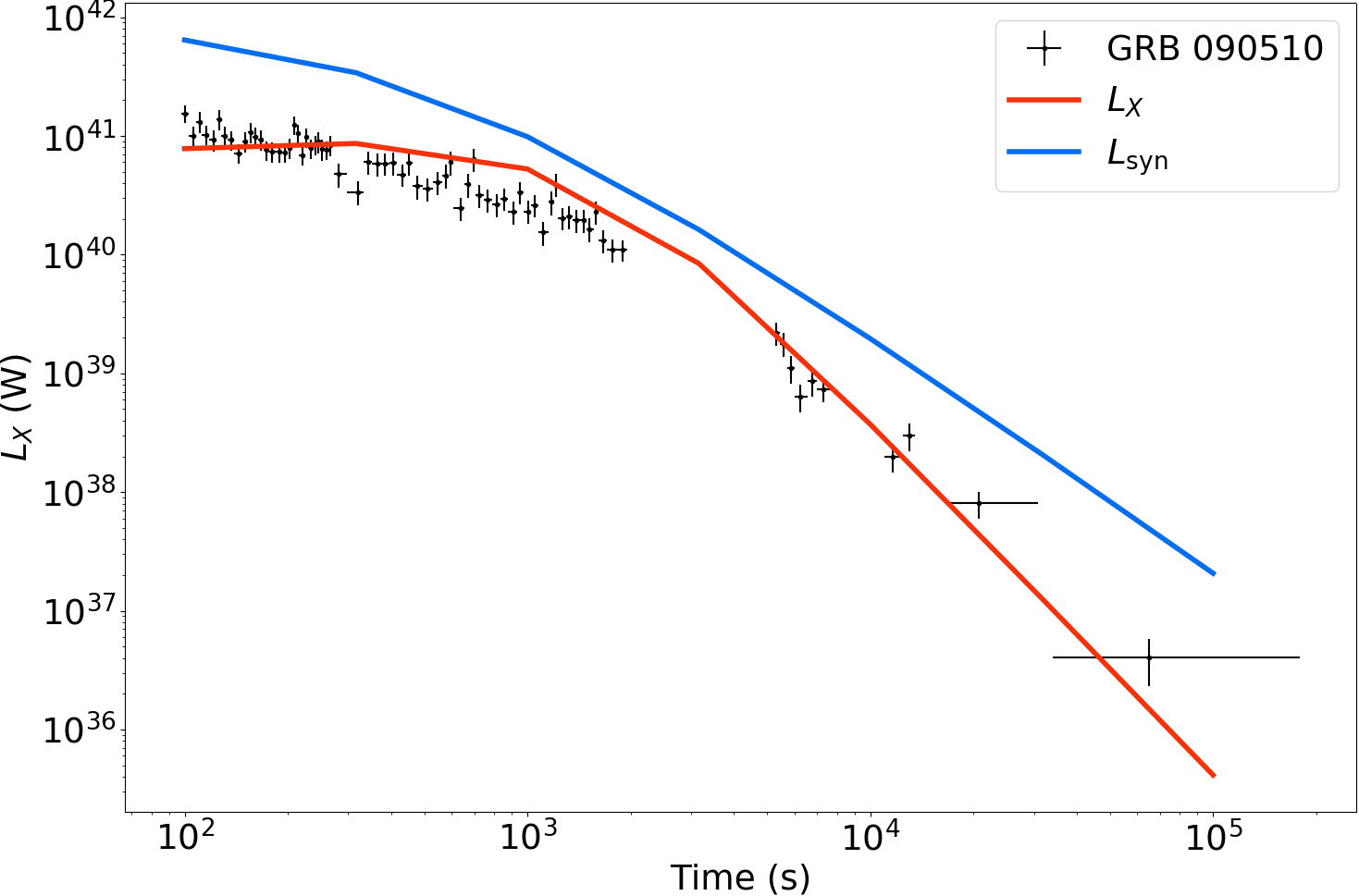}
  \caption{\label{fig:plateaudat} X-ray luminosity $L_X$ (red curve); and bolometric luminosity $L_{\rm syn}$ (blue curve) (in watts) versus time (in seconds). The curves are the model predictions for the fiducial parameters $B_0 = 5\times 10^{11}$ T, $\Omega_0 = 1.5 \times 10^3$ rad s$^{-1}$. The data are \textit{Swift} observations of GRB 090510. The solid curve is not a fit to the data but shares some of the qualitative features observed.}
\end{figure}

\begin{figure}
  \centering
  \includegraphics[width=\linewidth]{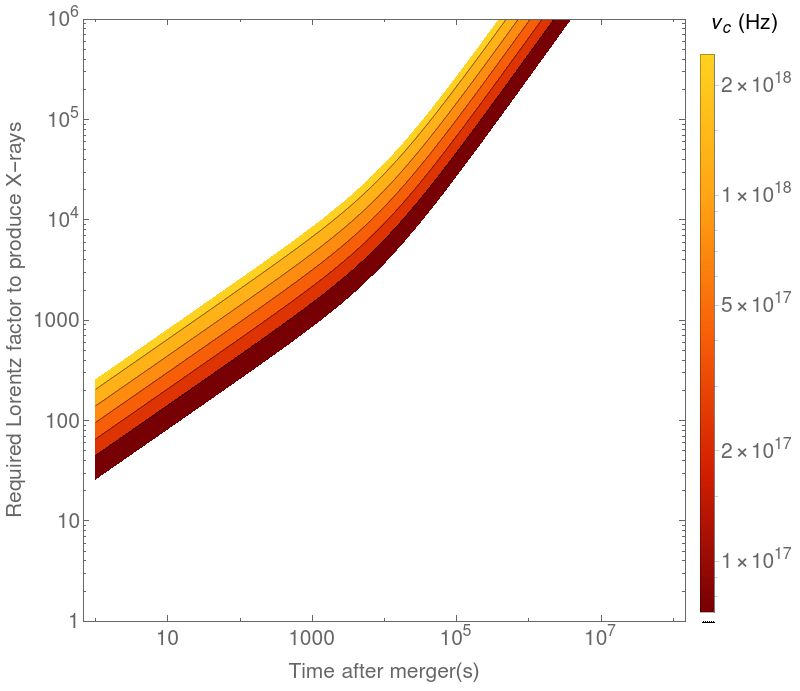}
  \caption{\label{fig:charfreq} Electron Lorentz factor $\gamma_e$ needed to emit in the X-ray band $0.3$ -- $10$ keV as a function of time (s) in a magnetic field that varies as (\ref{eqn:magfieldt}) with $B_0 = 10^{12}$ T and $\Omega_0 = 10^3$ rad s$^{-1}$. Each contour represents the characteristic frequency for an electron at time $t$ with energy $\gamma_e$ (see colour bar at right). The regions in white correspond to characteristic frequencies outside the X-ray band.}
\end{figure}

Figure \ref{fig:plateaudat} presents the X-ray light curve predicted by the full plerion model with $B(t)$ given by (\ref{eqn:magfieldt}) and adiabatic expansion. 
It contains two components: the plateau for $t \lesssim \tau$ and a rapid drop-off for $t \gtrsim \tau$.
The `plateau' is not flat in the X-ray band.
It rises as $L_X \propto t^{2}$ until it reaches a peak, after which it falls off as $L_X \propto t^{-2}$.
As the electron bubble expands, $B(t)$ decreases, and the electron energy required to generate X-ray radiation increases according to (\ref{eqn:charfreq}) as $E \propto t^{1/2}$.
Hence the number of electrons emitting in the X-ray band varies according to $B(t)$ and the injected power law.
In the second stage, for $t \gtrsim \tau$, the predicted X-ray flux drops more rapidly than the bolometric flux.
This is driven by $B(t)$ and hence $\nu_c$, which decrease as above, and the decreasing injection rate which scales as $\dot{N}_{\text{inj}}\propto t^{-2}$.
Figure \ref{fig:charfreq} shows the evolution of the electron Lorentz factor required to produce X-ray radiation with $7.25\times 10^{16} \leq \nu_c/({\rm Hz }) \leq 2.42\times 10^{18}$.

The features predicted by the model align with observations in a broad, qualitative sense.
In Figure \ref{fig:plateaudat}, we overplot \textit{Swift} measurements of $L_X$ for GRB 090510 \citep{Evans2007,Evans2009} scaled to a redshift of $z = 0.9$ \citep{2009GCN..9353....1R}.
We stress that this is not a fit; we are merely comparing observational features qualitatively.
Nevertheless, the shapes of the predicted and measured light curves are similar for both $t \lesssim \tau$ and $t \gtrsim \tau$, and the normalization agrees within an order of magnitude despite the idealized nature of the model.

An X-ray plateau also emerges from the simpler model discussed in Appendix \ref{sec:sol}, with $B(t) = B = $ constant and no adiabatic expansion.
This plateau is flat and satisfies $L_{\rm syn}(t) \approx L(t)$, when the synchrotron loss time is short ($\lesssim 1$ s) for $c_s^{-1} \lesssim E B^2$.
Below this threshold, $L_X(t)$ may rise, until the cooling rate exceeds the energy injection rate. 

Both the simple and complete models feature a plateau phase for $t < \tau$.
They enter a second phase at $t > \tau$, as the injected luminosity begins to fall off. 
The rate of decline in the X-ray band is steeper than in the bolometric light curve, because the decline is driven by both the spin-down luminosity of the pulsar $L \propto (1+t/\tau)^{-2}$ and $ B(t) \propto t^{-1} (1+t/\tau)^{-1}$.

\subsection{Decay time-scale}
\label{sec:timescale}

Observations indicate an anti-correlation between plateau luminosity and duration in both long and short GRBs \citep{Dainotti2010,Rowlinson2013}.
The latter reference identifies the plateau luminosity $L_{\rm plat}$ in the $1$-- $10^4$ keV band and  the plateau duration $T_{\text{plat}}$ from \textit{Swift} light curves and reports
\begin{equation}
L_{\rm plat} = 10^{48.75 \pm 0.55} (T_{\rm plat}/{1 \, {\rm s}})^{-1.29\pm 0.12} \, {\rm W}.
\label{eqn:platrel}
\end{equation}
In this result $T_{\rm plat}$ is taken to be from $t = 0$ to the endpoint of the flat, middle segment in a three-segment, piecewise-power-law fit.\footnote{Equivalently, ``from the initial formation of the magnetar (i.e. the time of the GRB) and the point at which the X-ray emission from the magnetar starts to turn over from the plateau phase to a power-law decay phase'' (verbatim quote) \citep{Rowlinson2013}.}

\begin{figure}
  \centering
  \includegraphics[width=\linewidth]{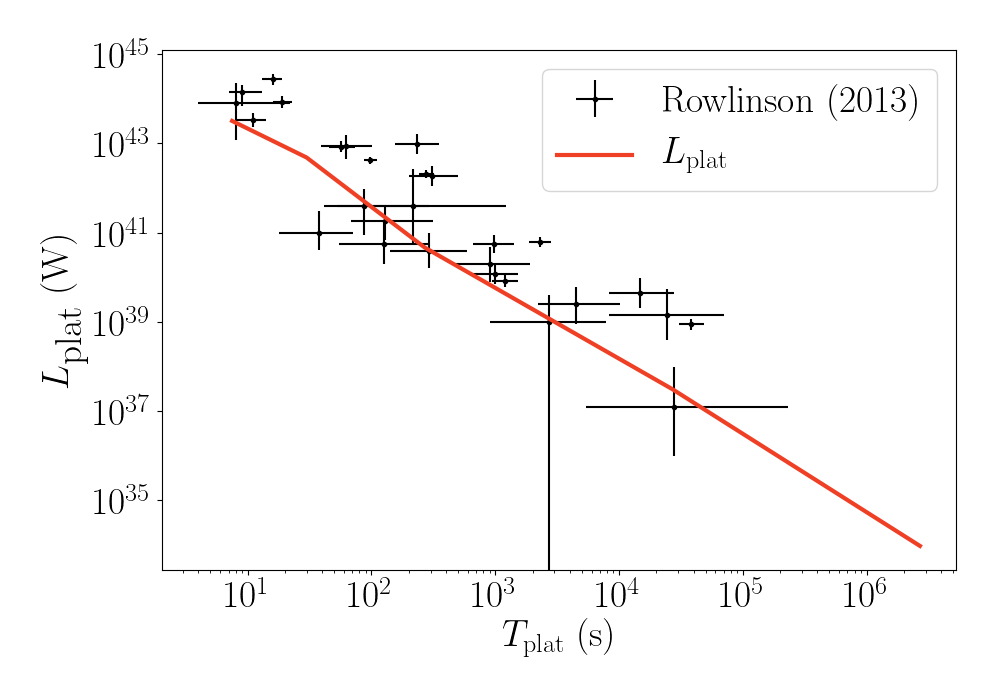}
  \caption{\label{fig:plateautimescales} Plateau luminosity $L_{\rm plat}$ (W) versus $T_{\rm plat}$ (s). Black points are observations reported by \citet{Rowlinson2013}. The solid red curve shows the model prediction $L_{\rm plat} = L_X(t=\tau)$ ($1$-- $10^4$ keV band) versus $T_{\rm plat} = \tau$ for varying $B_0$ and $\Omega_0$.}
\end{figure}

Does the plerion model in Sections \ref{sec:plerionmodel} and \ref{sec:spectralevolution} reproduce the observed plateau flux-duration anti-correlation?
In order to compare to \citet{Rowlinson2013}'s data, in this section only we work in the range $1$-- $10^4$ keV instead of $0.3$ -- $10$ keV.
The plateaux in our model are not perfectly flat, so we define $T_\mathrm{plat} = \tau$ and $L_{\rm plat} = L_X (t = \tau)$.
There are other valid definitions for $T_\mathrm{plat}$ and $L_{\rm plat}$ which give similar results.
We generate $L_\mathrm{X}(t = \tau)$ for the parameter combinations in Table \ref{tab:params} and plot the results against $T_\mathrm{plat} = \tau$ in Figure \ref{fig:plateautimescales} (red curve), together with the 26 data points taken from \citet{Rowlinson2013} (black crosses). 
Light curves with shorter $T_{\rm plat}$ (higher $B_0$) exhibit brighter plateaux in the model, and the X-ray luminosities are comparable to those observed.

\begin{table}
  \centering
\begin{tabular}{llll}
  \hline
\(B_0\) (T) & \(\Omega_0\) ( rad s\(^{-1}\) ) & \(L_0\) (W) & \(\tau\) (s)\\
\hline
\(10^9\) & \(1\times 10^3\) & \(1.86\times 10^{ 35 }\) & \(2.69\times 10^8\)\\
\(10^{10}\) & \(1\times 10^3\) & \(1.86\times 10^{37}\) & \(2.69\times 10^6\)\\
\(10^{11}\) & \(1\times 10^3\) & \(1.86 \times 10^{39}\) & \(2.69\times 10^4\)\\
\(10^{12}\) & \(1\times 10^3\) & \(1.86 \times 10^{41}\) & \(2.69 \times 10^2\)\\
\(10^{12}\) & \(3\times 10^3\) & \(1.50 \times 10^{43}\) & \(2.99 \times 10\)\\
\(10^{12}\) & \(6\times 10^3\) & \(2.41\times 10^{44}\) & \(7.48\)\\
  \hline
\end{tabular}
\caption{Combinations of $B_0$ and $\Omega_0$ used to generate the model predictions in Figure \ref{fig:plateautimescales}.}
\label{tab:params}
\end{table}

\subsection{Black hole formation}
\label{sec:bhform}

If the neutron star formed by the merger is supermassive, it may collapse at time $t_c$ to a black hole once it loses sufficient rotational energy.
It is sometimes argued that the X-ray flux turns off suddenly when this occurs.
However, if a magnetic field persists, the plerion continues to emit synchrotron radiation for a while, even when injection from the central engine ceases.

Once the injection ceases, $N(E,t)$ evolves according to the homogeneous solution to (\ref{eqn:pdegen}), with $N(E, t=t_c)$ determined by the pre-collapse evolution, and the post-collapse magnetic field structure $B_\mathrm{BH}(t)$.

\begin{figure}
  \centering
  \begin{subfigure}{ \linewidth }
  \includegraphics[width=\linewidth]{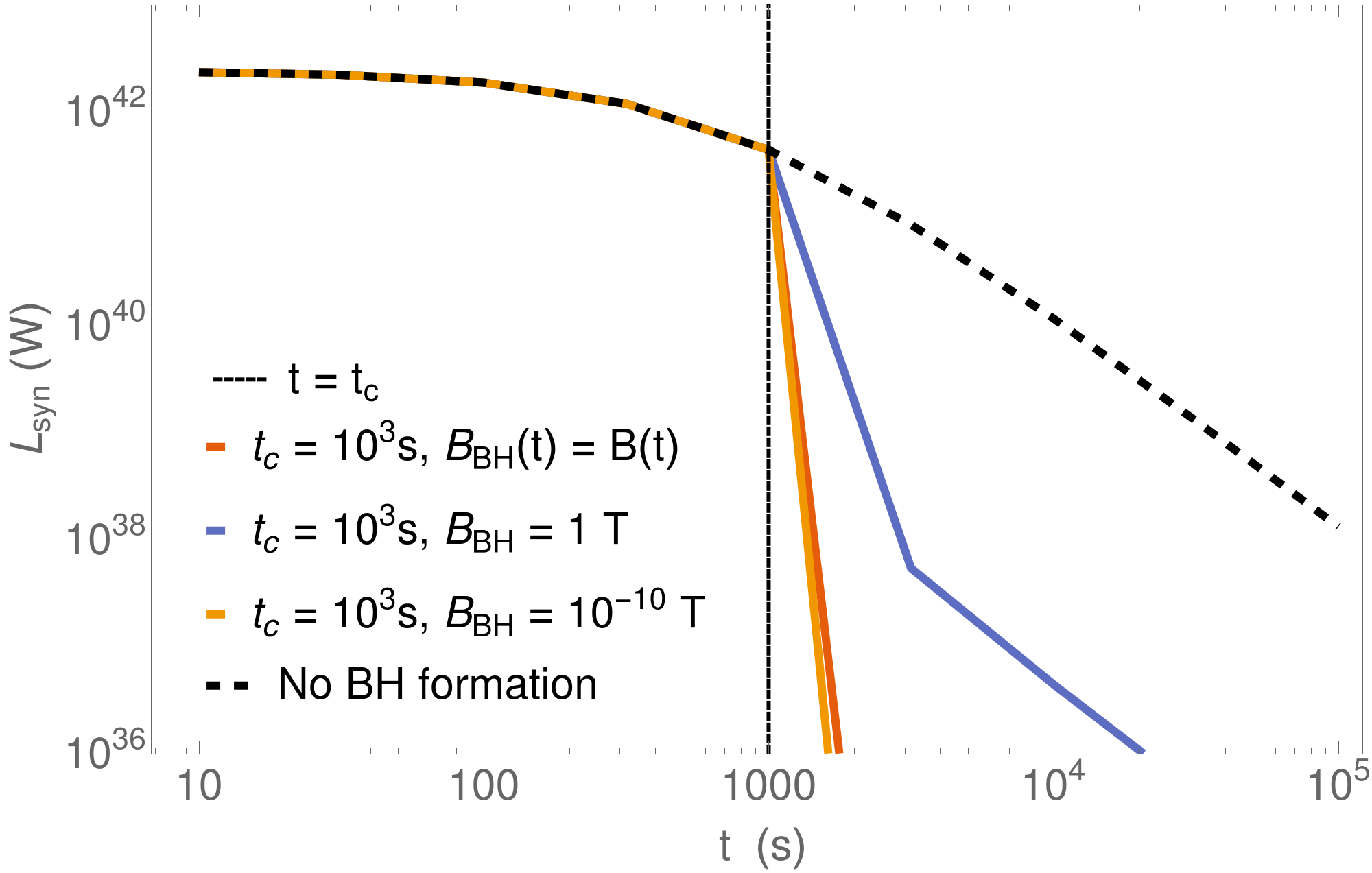}
  \caption{\label{fig:bhform} }
  \end{subfigure}
  \begin{subfigure}{ \linewidth }
  \includegraphics[width=\linewidth]{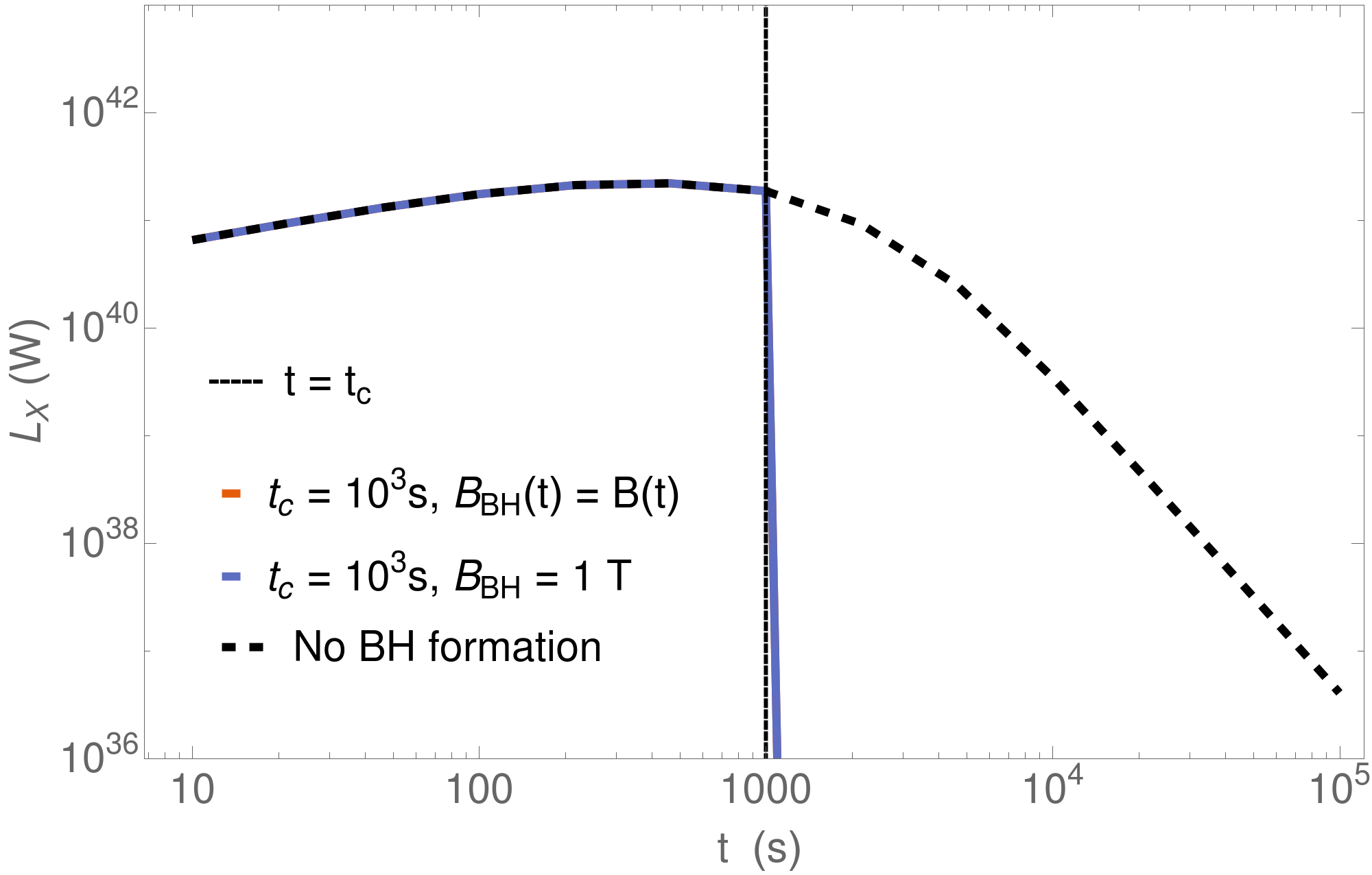}
  \caption{\label{fig:bhxray} }
  \end{subfigure}
  \caption{(a) Bolometric luminosity $L_{\rm syn} \,({\rm W})$ and (b) X-ray luminosity $L_X \, ({\rm W})$ versus time $t$ (s) when the magnetar collapses to a black hole at $t_{\rm c } = 10^3$ s (vertical line). Magnetic field scenarios: $B_{\rm BH}(t)$ equals a constant ISM value (yellow curve), $B(t)$ in the absence of collapse (red curve), and an intermediate value (blue curve). The dashed line shows the evolution of the remnant if the magnetar is stable and does not collapse to a black hole. The post-collapse emission is predominantly below the X-ray band.}
  \label{fig:bh}
\end{figure}

The synchrotron cooling time for an electron radiating at $\nu_{\rm c}$ is given by
\begin{equation}
  t_{\rm cool} = 3 \times 10^{-3} {\rm s} \left(\frac{B_{\rm BH}}{\rm 1 T} \right)^{-3/2} \left(\frac{\nu_c}{\rm 10^{17} Hz}\right)^{-1/2}
  \label{eqn:synchtime}
\end{equation}
from (\ref{eqn:synchdef}) and (\ref{eqn:charfreq}), where $B_{\rm BH}(t)$ is the characteristic, post-collapse magnetic field at $r\approx r_{\rm b}$.
Figure \ref{fig:bh} displays the light curve following black hole formation in two astrophysical scenarios. 
In the first scenario, the magnetic field threading the black hole magnetosphere is maintained at the same strength as the pre-collapse $B_0$, and the magnetic field in the plerion (at $r\approx r_{\rm b}$) equals $B(t)$ in the absence of collapse.
In this scenario (Figure \ref{fig:bhform}, yellow curve), the plerion continues to emit synchrotron radiation bolometrically for $t \gg t_{\rm c }$, but $L_X$ cuts off at $t \approx t_{\rm c }$ (Figure \ref{fig:bhxray}) because no more high energy electrons are injected, and the existing population radiates its energy on a timescale $\lesssim 1$ s from ~(\ref{eqn:synchtime}).
Emission is therefore predominantly at lower frequencies.
In the second scenario, the black hole does not maintain the pre-collapse magnetic field, and $B_{\rm BH} \sim 10^{-10}\,{\rm T}$ reverts to a value typical of the ISM (Figure \ref{fig:bhform}, blue curve).
In this case, the luminosity cuts off at all frequencies because $B_{\rm BH}$ is too weak to produce synchrotron radiation bright enough to be observed.

\subsection{Remnant properties at birth}
\label{sec:birth}
Observational properties of the light curve can be used to infer properties of the neutron star in principle.
The light curve generated by the model is sensitive to $B_0$ and $\Omega_0$.
As discussed in Section \ref{sec:timescale}, $\tau$ affects not only the turnover point in the light curve but also its brightness.

\begin{figure*}
  \centering
  \begin{subfigure}{0.45\linewidth}
  \includegraphics[width=\linewidth]{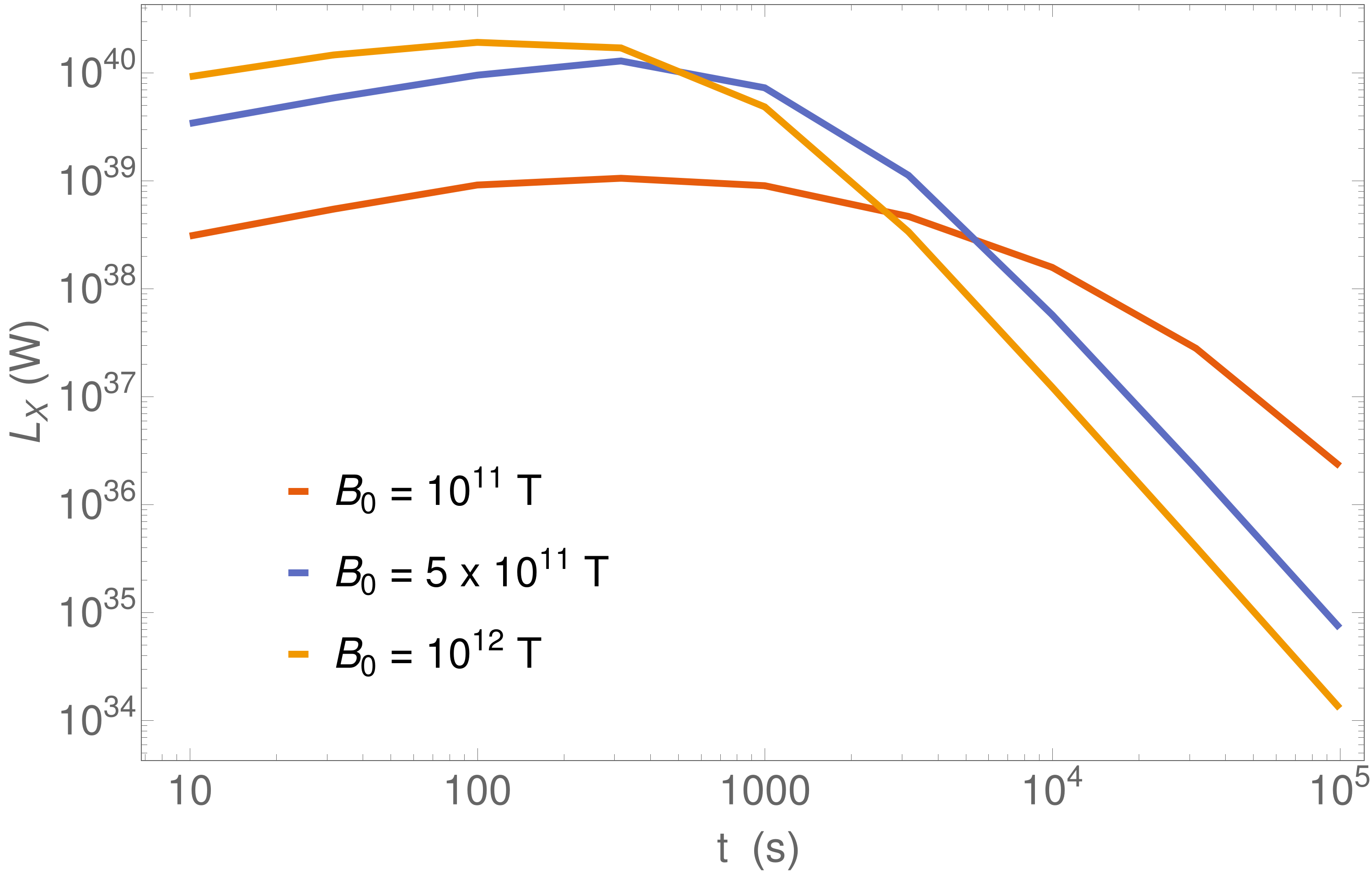}
  \caption{\label{fig:bcomp}}
  \end{subfigure}
  \begin{subfigure}{0.45\linewidth}
  \includegraphics[width=\linewidth]{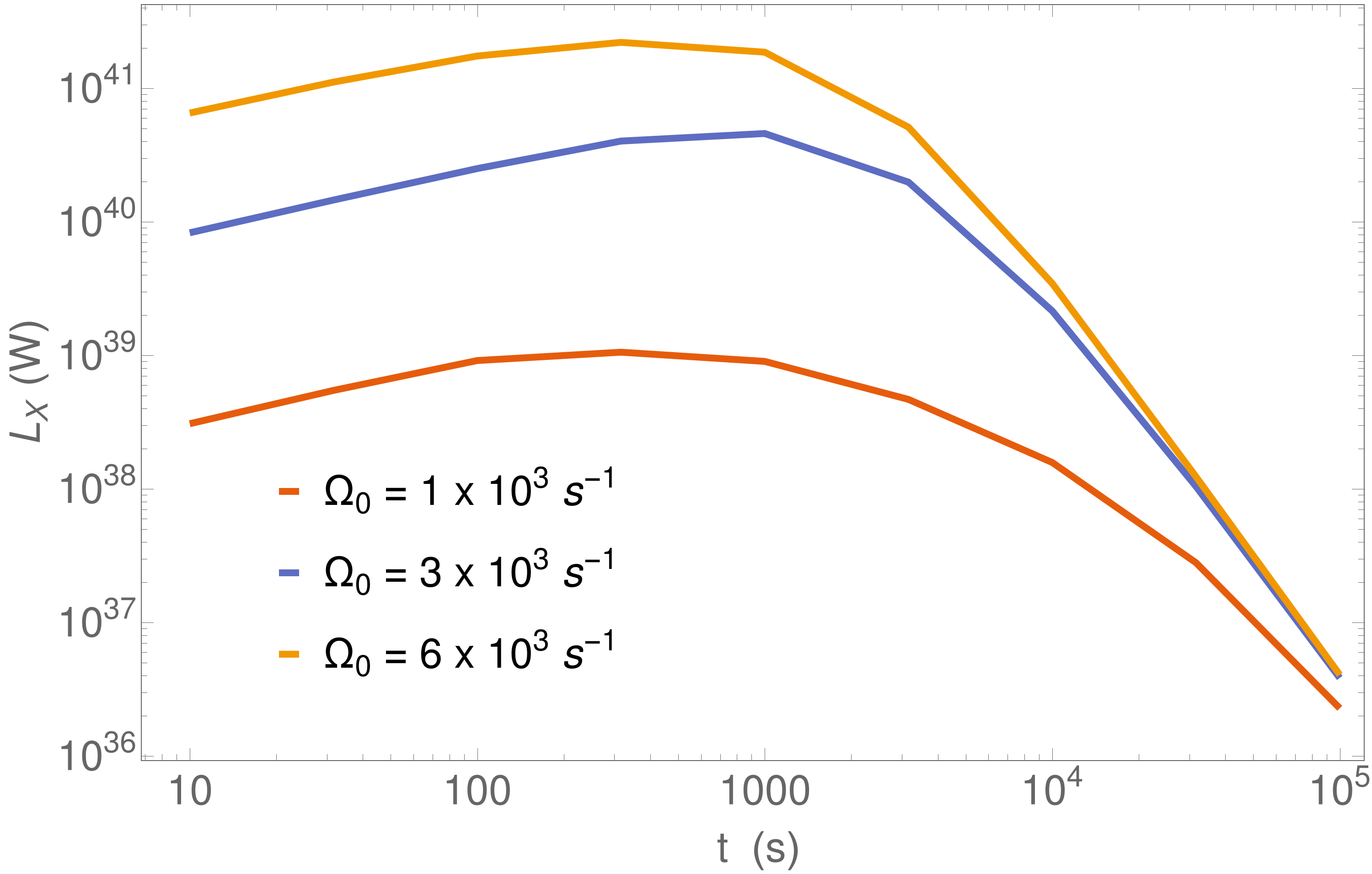}
  \caption{\label{fig:spincomp}}
  \end{subfigure}
  \begin{subfigure}{0.45\linewidth}
  \includegraphics[width=\linewidth]{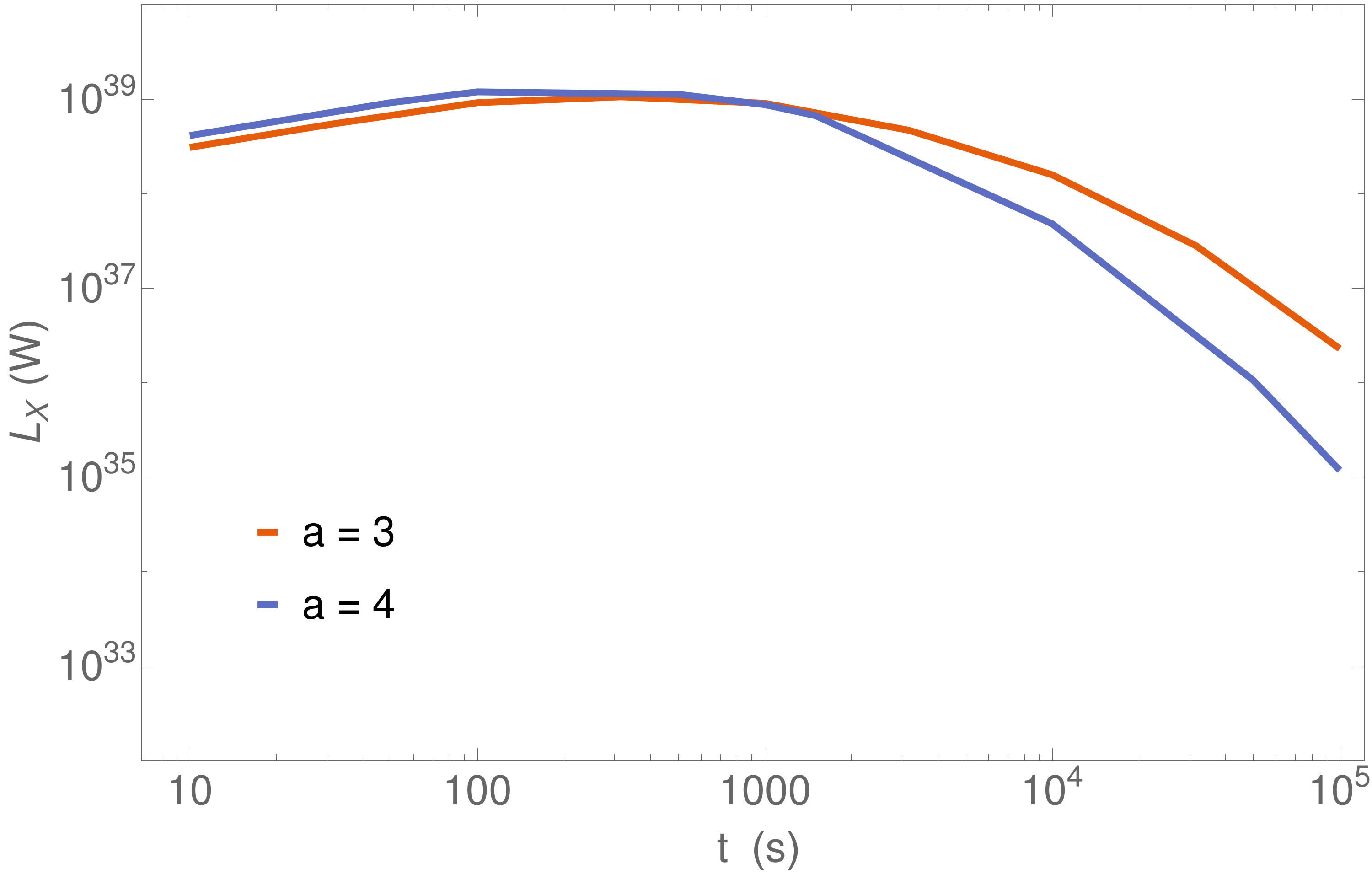}
  \caption{\label{fig:acomp}}
  \end{subfigure}
  \begin{subfigure}{0.45\linewidth}
  \includegraphics[width=\linewidth]{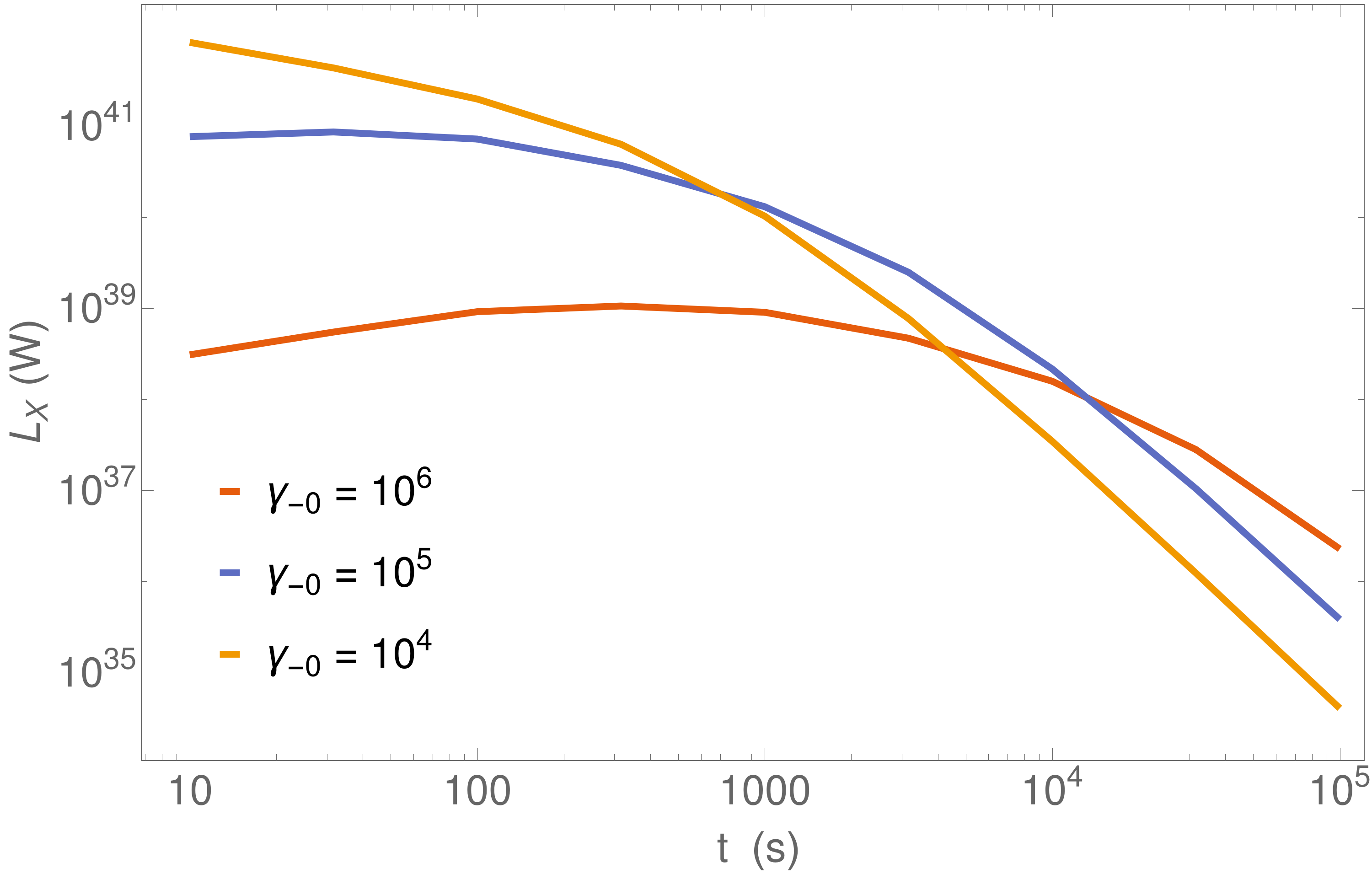}
  \caption{\label{fig:ecomp}}
  \end{subfigure}
  \caption{Predicted X-ray luminosity (W) versus time (s) for varying (a) surface magnetic field $B_0$, (b) initial angular velocity $\Omega_0$, (c) injection exponent $a$, and (d) minimum injection energy $E_{-0}$. Curves are colour coded according to the legends. Fixed parameters: (a) $\Omega_0 = 10^3 \, {\rm rad \, s}^{-1}$, $a = 3$, $E_{-0} = 8\times 10^{-11} {\rm J}$; (b) $B_0 = 10^{11} \, {\rm T}$, $a = 3$, $E_{-0} = 8\times 10^{-11} {\rm J}$; (c) $B_0 = 10^{11} \, {\rm T}$, $\Omega_0 = 10^3 \, {\rm rad \, s}^{-1}$, $E_{-0} = 8\times 10^{-11} {\rm J}$; (d) $B_0 = 10^{11} \, {\rm T}$, $\Omega_0 = 10^3 \, {\rm rad \, s}^{-1}$, $a = 3$.}
  \label{fig:comp}
\end{figure*}

The polar magnetic field $B_0$ influences both the plateau luminosity and its duration.
A stronger magnetic field corresponds to a brighter plateau.
In Figure \ref{fig:bcomp}, raising the magnetic field from $10^{11}$ T (red curve) to $10^{12}$ T (yellow curve) increases the plateau luminosity by two orders of magnitude.
The duration of the plateau shrinks accordingly, producing a sharper plateau turn-off.

The overall luminosity of the plateau is affected by $\Omega_0$.
In Figure \ref{fig:spincomp}, increasing $\Omega_0$ from $1\times 10^3$ rad s$^{-1}$ to $6\times 10^3$ rad s$^{-1}$ raises the luminosity of the plateau by three orders of magnitude and decreases the plateau duration slightly.
This does not contradict Figure \ref{fig:plateautimescales}, because the calculations in Section \ref{sec:timescale} refer to the $1$ -- $10^4$ keV band, as opposed to the $0.3$ -- $10$ keV band used throughout the rest of the paper.
Furthermore, $\Omega_0$ and $B_0$ are varied jointly in Figure \ref{fig:plateautimescales}, whereas only $\Omega_0$ varies in Figure \ref{fig:spincomp}.

In addition to $B_0$ and $\Omega_0$, the shape of the curve is affected by the minimum injection energy $E_{-0}$, and the injection index $a$ in $\dot{N}_{\rm inj}(E, t) \propto E^{-a}$. 
As discussed in Section \ref{sec:flux}, the predicted flux decreases as the magnetic field decreases, and the number of electrons emitting at X-ray frequencies therefore diminishes.
This effect is subdominant compared to other mechanisms until
\begin{equation}
B(t) \lesssim \frac{4\pi m_e (m_e c^2)^2 \nu_\mathrm{c}}{3 E_{-0}^2 e} .
\label{eqn:nubreak}
\end{equation}
Hence the brightness in the X-ray band increases when $E_{-0}$ increases.
This can be seen in Figure \ref{fig:ecomp}, which compares the light curves for $E_{-0} = 10^{-8}$ J and $10^{-12}$ J.
Similarly, a higher $a$ value produces a shorter plateau in the X-ray band, as a steeper power law contains fewer X-ray emitting electrons at high energies.
Note that the slope of the plateau is insensitive to $a$ in the range $3 \leq a \leq 4$.

\section{Merger Ejecta}
\label{sec:ejecta}

\begin{figure}
  \centering
  \includegraphics[width=\columnwidth]{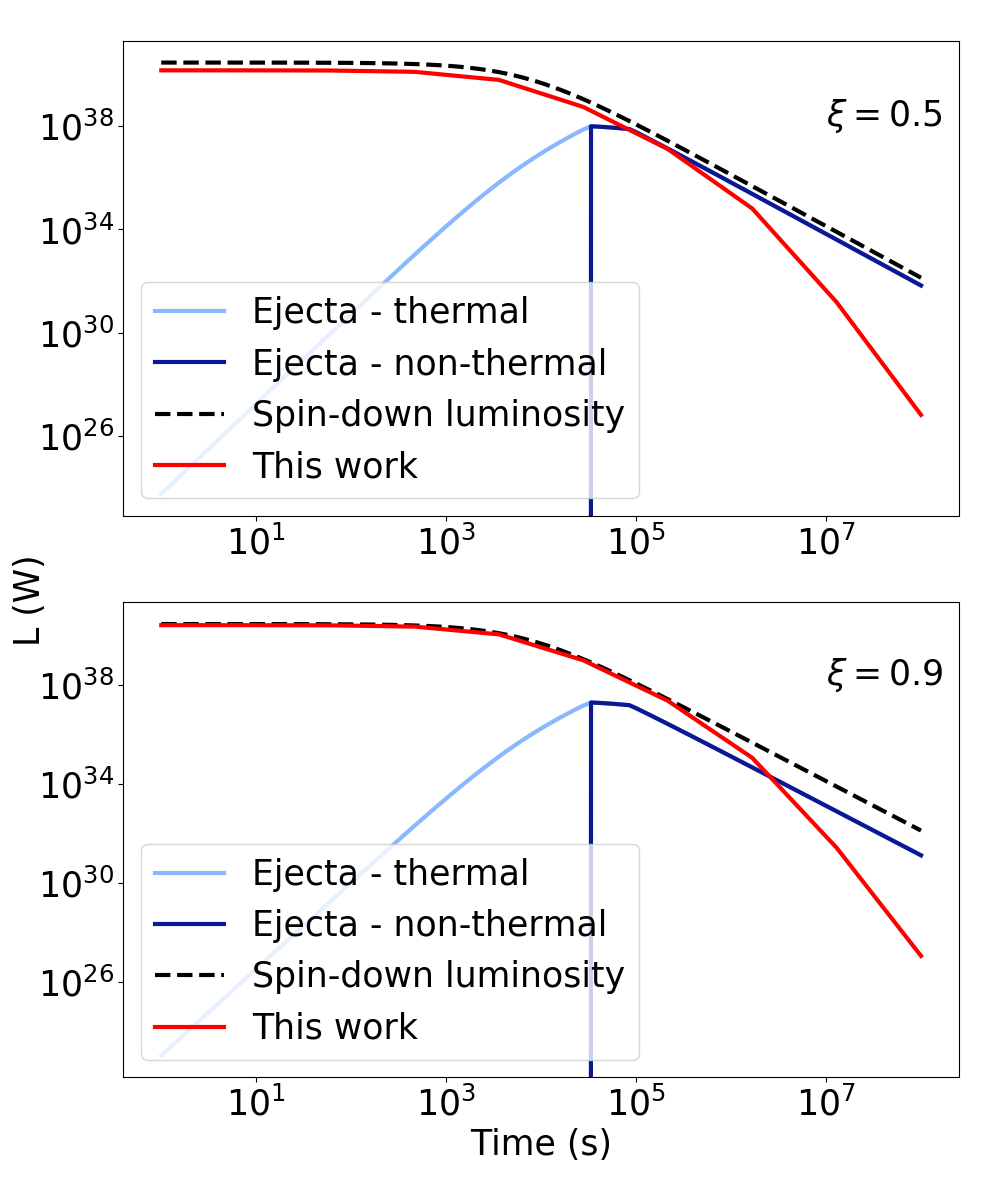}
  \caption{\label{fig:ejecta} Predicted bolometric luminosity (W) versus time (s) from the plerion (Sections \ref{sec:plerionmodel} -- \ref{sec:xray}) and the merger ejecta \citep{Metzger2014} for (a) $\xi = 0.5$ and (b) $\xi = 0.9$, where $\xi$ is the fraction of the spin-down luminosity fuelling the plerion model. The thermal (optical/UV) and non-thermal ejecta (X-ray) are shown in light and dark blue respectively. The plerionic flux is shown in red. The black dashed curve is the spin-down luminosity of the magnetar.}
\end{figure}

In Sections \ref{sec:plerionmodel}--~\ref{sec:xray} we assume for clarity that the plerionic emission is not absorbed by the merger ejecta. 
Thus far, we make the simplifying assumption that the ejecta do not affect the evolution of the remnant.
Previous authors \citep{Yu2013,Metzger2014,Siegel2016} studied carefully the opposite extreme, where the  magnetar is completely shrouded by the ejecta, and the plerionic emission is reprocessed to produce an optically thick non-thermal spectrum.
In reality an sGRB is likely to lie between the two extremes.
While there is certainly a surrounding shell of material as observed in GW170817 \citep{Cowperthwaite2017,Tanaka2017,Evans2017}, that material may be pierced by a jet, perforated by Rayleigh-Taylor instabilities, or otherwise disturbed \citep{Arons2003}.
In this scenario, a geometry dependent-fraction of the plerionic emission is intercepted by the ejecta and reprocessed, while the remainder escapes.
The light curve of the observed transient is the sum of the components.
Here we parameterize the situation crudely by assuming that a fraction $\xi$ of the pulsar spin-down luminosity $L(t)$ drives the plerion, which emits as in Sections \ref{sec:plerionmodel}--~\ref{sec:xray}, while the rest drives the ejecta, which emit according to the analytic theory in the appendix of \citet{Metzger2014}.
This approach is adequate for gaining an approximate sense of the relative contributions of the two components.
However, it is not a substitute for a self-consistent calculation, which lies outside the scope of this paper and should include multiple refinements introduced by previous authors \citep{Yu2013,Metzger2014,Siegel2016}.

In the simplest scenario, the radiation from the ejecta consists of two components: an optical, thermal transient driven by the absorption and re-emission of X-rays by the ejecta, and a late-time X-ray transient that appears only after the ejecta are fully ionised \citep{Metzger2014}.
These components are the thermal and non-thermal curves in Figure \ref{fig:ejecta}.
The optical thermal component dominantes for $t < \tau$ but is dimmer than the plateau predicted by our model.

Figure \ref{fig:ejecta} compares the predicted light curves from \citet{Metzger2014} with the plerionic emission from a magnetar with $B_0 = 10^{11} {\rm \, T}$ and $\Omega_0 = 2\times 10^3 {\rm \, rad \, s}^{-1}$.
At $t > \tau$, the non-thermal ejecta component becomes the dominant source of luminosity.
It outshines the plerionic emission at late times. 
The plerionic flux declines not just due to the $t^{-2}$ scaling of $\xi L(t)$, but also due to the magnetic field declining as $t^{-2}$ for $t > \tau$ (as discussed in Section \ref{sec:flux}).
In contrast, the non-thermal flux from the ejecta scales as $(1-\xi) L(t) \propto t^{-2}$ and therefore dominates late times.
This can be seen Figure \ref{fig:ejecta}: regardless of $\xi$, the plerionic flux always ends up dimmer than the non-thermal ejecta flux.
Part of this result may flow from the simple approach we take to combining the two components.
The non-thermal emission in \citet{Metzger2014} becomes visible, as the expanding (and possibly ionizing) ejecta evolve.
The analytic light curves from \citet{Metzger2014} are constructed assuming that the spin-down luminosity of the magnetar directly fuels the ejecta.
In a composite model, the X-rays incident onto the ejecta are part of the plerionic spectrum, which in turn affects the shape of the reprocessed spectrum.

Strictly speaking, a proper treatment of the sGRB remnant involves integrating the plerionic and ejecta at a fundamental level and within a realistic three-dimensional geometry..
Building a hybrid model from the ground up is a priority for future work.
The value of the preliminary order-of-magnitude estimate in this section is that it gives a sense of when the plerion and ejecta dominate the light curve as a function of $\xi$.

\section{Discussion}
\label{sec:discussion}
Millisecond magnetar models explain some observed features of canonical sGRB afterglows. 
The work presented here provides some insight into the possible physics behind the observed plateaux in X-ray light curves for sGRBs.
By calculating the spectral evolution of the electron population $N(E,t)$ due to synchrotron radiation, adiabatic cooling, and power-law injection, we are able to reproduce certain observed features in canonical X-ray light curves.
The model generates X-ray plateaux with comparable luminosities and durations to those observed. 
The plateaux are not flat; the flux is predicted to decrease slowly with time, in accord with what is observed.
The X-ray luminosity is sensitive to the energy distribution of the injected electrons, in particular the minimum injection energy $E_{-0}$ and the power-law index $a$.
An anti-correlation between plateau luminosity and duration is a natural consequence of the model.
We also find that the plerionic emission can dominate the X-ray light curve at early times and remains brighter than the non-thermal emission from the ejecta, depending on the three-dimensional geometry and hence fraction of the plerionic emission that is reprocessed.

We deliberately neglect many important details in this idealized model, such as spatial variation, diffusion of energetic electrons in position and energy, details of the shock structure (e.g. anisotropy) and relativistic beaming.
We over-simplify the treatment of reproessing by the merger ejecta, as noted above.
We also neglect observational features which deviate from a `canonical' lightcurve such as late-time X-ray flares or extended emission \citep{Gompertz2013}.
These features may be incorporated into the plerion model with little mathematical difficulty given the Green function framework in this paper.
For example, X-ray flares may be produced by multiple injection events.
We leave these interesting questions for future work.
The intent of this paper is not to provide a complete picture of the remnant but rather to generalize existing magnetar models to track the evolution of $N(E,t)$ within the plerionic component.
Future modelling which improves on the above approximations should be compared to a broad sample of sGRBs.

\section*{Acknowledgements}
We thank the anonymous referree for drawing our attention to the importance of the merger ejecta and several key references.
Parts of this research were conducted by the Australian Research Council Centre of Excellence for Gravitational Wave Discovery (OzGrav), through Project Number CE170100004.
The work is also supported by an Australian Research Council Discovery Project grant (DP170103625).
This work makes use of data supplied by the UK Swift Science Data Centre at the University of Leicester.



\bibliographystyle{mnras}
\bibliography{xray} 



\appendix

\section{Analytic solution for $N(E,\lowercase{t})$ with constant injection and magnetic field}
\label{sec:sol}
In this appendix, we examine simplified versions of the model, in which the magnetic field and electron injection are constant.
These versions may be solved analytically, demonstrating the mathematical technique behind the X-ray light curves for the full, unsimplified model in Section \ref{sec:xray}.

\subsection{Green's function}
\label{sec:green}
In order to solve (\ref{eqn:pdegen}) for an arbitrary, time-dependent, power-law source $S(t)$, we first solve for the impulse response (Green's function) $G(E, t; t_i)$, which satisfies
\begin{equation}
  \frac{\partial G}{\partial t} - \frac{\partial  }{\partial E}\left[\frac{dE}{dt}_{\rm cool} G(E,t;t_i) \right] = E^{-a} \delta(t - t_i),
\label{eqn:greenspde}
\end{equation}
i.e. instantaneous power-law injection at $t = t_i$.
We then integrate the Green's function weighted by the source to obtain

\begin{equation}
N(E,t) = \int dt_i G(E,t;t_i) S(t_i).
\label{eqn:greens}
\end{equation}
Integrating (\ref{eqn:greenspde})  with $G(E, t; t_i) = 0$ for $t < t_i$, we find the initial condition
\begin{equation}
G(E, t_i; t_i) = E^{-a}.
\label{eqn:initcond}
\end{equation}

For the case of a constant magnetic field $B$, and neglecting adiabatic cooling temporarily,  the homogeneous solution is
\begin{equation}
N(E,t) = E^{-2} f \left( \frac{c_s B^2 E t - 1}{c_s B^2 E}\right) 
\end{equation}
where $f(x)$ is a function determined by the boundary conditions.
Using (\ref{eqn:initcond}), we find 
\begin{equation}
  G(E, t;t_i) = E^{-2} \left[ c_s B^2 (t_i -t)+E^{-1}\right]^{a-2}.
  \label{eqn:greensconst}
\end{equation}
The case $a=2$ is a curious one, both physically and mathematically.
The solution reduces to the injection law $G(E, t; t_i) = E^{-2}$, implying that the shape of the energy spectrum does not change, even though the maximum and minimum bounds decrease.
Physically this occurs because injection replenishes the electron population at an energy-dependent rate exactly equal to the depletion rate due to synchrotorn radiation.

\subsection{Energy range}
\label{sec:range}
There are three restrictions on the allowed energy range at any given $t$ and $t_i$. 
First, the Green's function must be real and positive.
For the case of a constant magnetic field without adiabatic cooling, this is equivalent to the requirement 
\begin{equation}
c_s B^2 (t -t_i) \geq E^{-1}
\label{eqn:gboundconst}
\end{equation}
for any time $t$ after the burst.
Second and third, the minimum and maximum electron energies evolve according to
\begin{equation}
E(t, t_i, E_{\pm}) = \left[E_{\pm 0}^{-1} + c_s B^2 (t-t_i)\right]^{-1}
\label{eqn:eboundconst}
\end{equation}
for a particular $t$, with $t_i \leq t$.
This does not define the minimum and maximum energies in the plerion as a whole.
Rather, it defines the maximum and minimum energy for a population of electrons injected at time $t_i$ in the range $E_{-0} \leq E \leq E_{+0}$.
So long as injection is constant, (\ref{eqn:gboundconst}) is satisfied identically when  (\ref{eqn:eboundconst}) holds.
This effectively shrinks the energy domain as time passes because the electrons must satisfy both these relations.

\begin{figure}
  \centering
  \includegraphics[width=\linewidth]{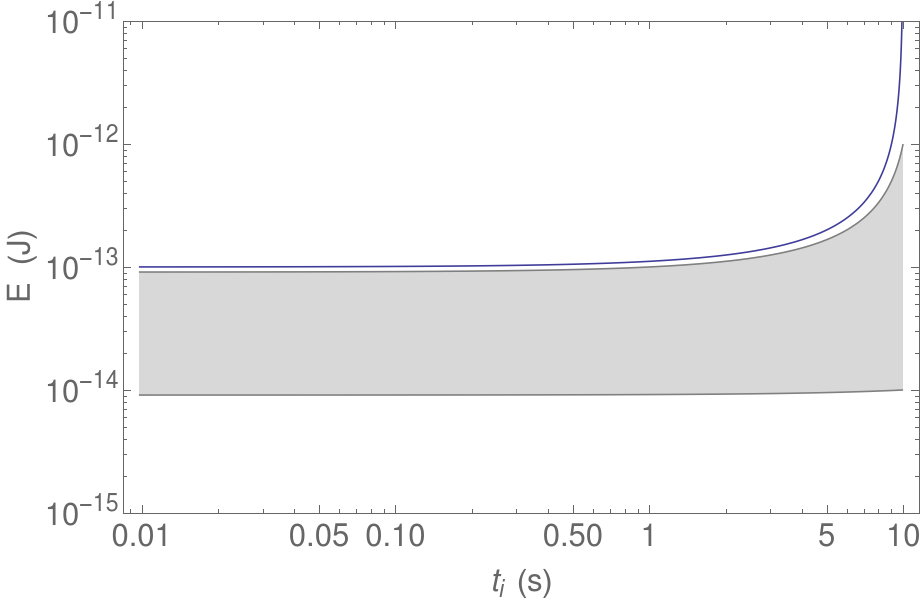}
  \caption{\label{fig:erange10} Allowed energy range on the $E-t_i$ plane for fixed $t = 10$ s. The navy curve is the maximum allowable energy for a given $t_i$ for a real and positive Green's function [equation (\ref{eqn:gboundconst})]. The grey shaded area is the allowable energy range defined by $(dE/dt)_{\mathrm syn}$ [equation (\ref{eqn:eboundconst})]. }
\end{figure}

\subsection{Bolometric luminosity}
\label{sec:bolometric}
A simple yet physically interesting scenario is the case $B(t) = $ constant and $L(t) = $ constant without adiabatic cooling.
The bolometric power output is defined as

\begin{equation}
L_{ \text{syn} }(t) = \int dt_i \int dE c_s B(t)^2 E^2 G(E, t; t_i) S(t_i)
\label{eqn: powerdef}
\end{equation}
The integral in (\ref{eqn: powerdef}) is performed over the shaded region in Figure \ref{fig:erange10}.
We consider the case $a=3$ for definiteness, but the principle is of course the same for any $a \geq 2$.
The bolometric luminosity is 
\begin{equation}
L_{\rm syn} = \frac{c_s L_0 B^2 t}{E_{-0}^{-1} - E_{+0}^{-1}}\ln \frac{E_{+0} + c_s B^2 E_{+0} E_{-0} t}{E_{-0} + c_s B^2 E_{+0} E_{-0} t}
\label{eqn:simplesol}
\end{equation}
Equation (\ref{eqn:simplesol}) satisfies $L_{\text{syn}}(t) \leq L_0$ for all $t$. 
At early times, the luminosity climbs from zero to its maximum value $L_0$, at which point the synchrotron radiation exactly balances the input spin-down luminosity.
In the limit $t \rightarrow \infty$, we find $L_{\text{syn}}(t) \rightarrow L_0$.
\subsection{Time-varying injection and magnetic field}
\label{sec:timevar}

The methods in Sections \ref{sec:green}--~\ref{sec:bolometric} may be applied to any time-dependent magnetic field and injection history, with or without adiabatic expansion included.

As an example, we consider the case where the injection varies with time according to $S(t_i) \propto L(t_i) \propto (1+t_i/\tau)^{-2}$, using the same conditions as in Section \ref{sec:bolometric} and integrating as before.
We obtain
\begin{equation}
\begin{aligned}
  L_\text{syn}(t) & = \frac{c_s L_0 B^2 \tau}{E_{-0}^{-1} - E_{+0}^{-1}}\left\{ \frac{\tau}{t+\tau}\ln \frac{E_{-0}}{E_{+0}} + \ln \frac{E_{-0}^{-1} + c_s B^2 t}{E_{+0}^{-1} + c_s B^2 t}\right. \\
  & + \frac{c_s B^2 E_{+0}^{-1}\tau }{1+c_s B^2 E_{+0}^{-1} (t + \tau)} \ln \left[\left( 1 + c_s B^2 E_{+0}t \right) \left(1 + \frac{t}{\tau}\right) \right]  \\
  & + \frac{t + c_s B^2 E_{+0}t (t+\tau)}{\left[t+\tau\right] \left[1+c_s B^2 E_{+0}\left(t+\tau\right)\right]^2} \\
 &   +\frac{ c_s B^2 E_{+0} \tau (t+\tau)\ln \left[\left( 1 + c_s B^2 E_{+0}t\right)\left(1+t/\tau\right)\right]}{\left[t+\tau\right] \left[1+c_s B^2 E_{+0}\left(t+\tau\right)\right]^2} \\
  & - \frac{c_s B^2 E_{-0}^{-1}\tau }{1+c_s B^2 E_{-0}^{-1} (t + \tau)} \ln \left[\left( 1 + c_s B^2 E_{-0}t \right) \left(1 + \frac{t}{\tau}\right) \right]  \\
 & - \frac{t + c_s B^2 E_{-0}t (t+\tau) }{\left[t+\tau\right] \left[1+c_s B^2 E_{-0}\left(t+\tau\right)\right]^2} \\
 & - \left.\frac{c_s B^2 E_{-0} \tau (t+\tau)\ln \left[\left( 1 + c_s B^2 E_{-0}t\right)\left(1+t/\tau\right)\right]}{\left[t+\tau\right] \left[1+c_s B^2 E_{-0}\left(t+\tau\right)\right]^2} \right\}.
\end{aligned}
\label{eqn:powerfull}
\end{equation}
This model is sufficient to reproduce the plateau and drop off phase, and is the form underlying the results for a constant magnetic field in Section \ref{sec:xray}.

It is possible to repeat the above calculation for a time-varying magnetic field combined with adiabatic expansion.
One can solve the homogeneous form of (\ref{eqn:pdegen}) via the method of characteristics and derive an analytic expression for the Green's function with the help of a symbolic algebra package such as Wolfram Mathematica.
This is exactly what is done to generate the results in Section \ref{sec:xray}.
We do not present the Green's function here, as it is too lengthy to yield any physical insight.


\bsp	
\label{lastpage}
\end{document}